 \documentclass[final,3p,times]{elsarticle}

 \usepackage{graphics}

\usepackage{amssymb}
\usepackage{cases}

 \biboptions{comma,square,compress,numbers,sort}

\journal{Physics Letters B}

\begin{document}

\begin{frontmatter}



\title{Tensor perturbations of $f(R)$-branes}


\author{Yuan Zhong}
\author{Yu-Xiao Liu\corref{cor1}}
  \ead{liuyx@lzu.edu.cn}
  \cortext[cor1]{The corresponding author.}
 \author{Ke Yang}

\address{Institute of Theoretical Physics, Lanzhou University,\\
           Lanzhou 730000, People¡¯s Republic of China}

\begin{abstract}

We analyze the tensor perturbations of flat thick domain wall branes
in $f(R)$ gravity. Our results indicate that under the transverse
and traceless gauge, the metric perturbations decouple from the
perturbation of the scalar field. Besides, the perturbed equation
reduces to the familiar Klein-Gordon equation for massless spin-2
particles only when the bulk curvature is a constant or when
$f(R)=R$. As an application of our results, we consider the
possibility of localizing gravity on some flat thick branes. The
stability of these brane solutions is also shortly discussed.
\end{abstract}


\end{frontmatter}


\section{Introduction}
The trapping of various matter fields on four-dimensional domain
walls has been successfully realized by using both non-gravitational
or gravitational methods~\cite{Rubakov1983,Visser1985}. Recently,
domain walls in five-dimensional space-time (called the bulk)
attract renew attentions from the physical community after Randall
and Sundrum~\cite{Randall1999,Randall1999a} pointed out that
massless four-dimensional graviton can be realized on a thin wall
(called the brane) if the extra dimension is large and warped
properly. Randall and Sundrum assume that all the observable matter
fields are confined on the brane while gravitons can propagate in
the bulk. For this reason, their model is also known as the (thin)
braneworld scenario, or the RS model for short. The braneworld
scenario is very important, because it, to some extent, solves some
of the long existed problems such as the hierarchy problem, and the
cosmological constant problem, etc., for reviews
see~\cite{Rubakov2001,Maartens2010}.

However, the thin braneworld model suffers a drawback: the whole
theory is singular at the location of the brane, for example, the
bulk curvature diverges at there and a junction condition must be
introduced. Besides, in the original braneworld scenario we are not
clear how does our world (as a brane) formed. Such problems are
solved in the so called thick braneworld models, in which gravity
couples with a background scalar
field~\cite{Gremm2000a,Csaki2000a,DeWolfe2000a,Bazeia2009,Dzhunushaliev2010a}.
The formation of the domain wall is ascribed to the non-linear
property of the gravitational system.

In general relativity, four-dimensional gravity has been
successfully realized in some thick braneworld
models~\cite{Gremm2000a,Csaki2000a,DeWolfe2000a,Bazeia2009}. The
trapping of various kinds of matter fields on single or multi branes
are also discussed for both thin and thick
branes~\cite{Randjbar-Daemi2000,Randjbar-Daemi2000a,Kakushadze2000,Youm2000,Oda2001,Oda2001a,Ringeval2002,Gogberashvili2003,Koley2005,Gogberashvili2006a,Melfo2006,Silva-Marcos2007,Liu2008c,Liu2008a,Liu2008,Liu2009,Liu2009a,Guerrero2010,Liu2010,Liu2010a}.

However, for some reasons, we have to take the contributions from
the higher order curvature terms into consideration. On one hand,
the adding of higher order curvature terms might render general
relativity renormalizable~\cite{Stelle1977}. On the other hand,
higher-order curvature invariants also appear in the effective
low-energy gravitational action of more fundamental theories, such
as the string theory~\cite{Vilkovisky1992}. To prevent the theory
from the spin-2 ghosts, and the well-known Ostrogradski
instability~\cite{Woodard2007}, the higher order curvature terms are
usually introduced as the Gauss-Bonnet term or an arbitrary function
of the curvature, namely, $f(R)$. Both of these two modified gravity
theories were applied in discussing a wide range of issues in
cosmology and higher energy physics, for details,
see~\cite{Sotiriou2010,DeFelice2010,Nojiri2010} and references
therein.

It was shown that the introduce of the Gauss-Bonnet term usually
imposes no impact on the localization of gravity on the
brane~\cite{Corradini2000,Giovannini2001,Nojiri2001,Neupane2001,Germani2002,Lidsey2003,Davis2003,Cho2003,Sami2004,Barrab`es2005}.
However, in $f(R)$ gravity things are more complex, since the
dynamical equations are of fourth-order. Based on the fact that
$f(R)$ gravity is conformally equivalent to a second-order gravity
theory~\cite{Barrow1988}, some thin braneworld models have been
constructed in the lower order
frame~\cite{Parry2005,Bronnikov2007,Deruelle2008,Balcerzak2008} by
introducing a conformal transformation. We hope this method still
valid for the discussions of thick branes. However, as stated
in~\cite{Barvinsky2008}, the method used in
refs.~\cite{Parry2005,Bronnikov2007,Deruelle2008,Balcerzak2008}
would lead to ambiguous when a background scalar field is
introduced.

For this reason, the thick brane solutions usually were found
directly in the higher order
frame~\cite{Afonso2007,Dzhunushaliev2010}. In~\cite{Afonso2007},
with a background scalar field, the authors offered us some
analytical thick brane solutions in both constant curvature spaces
and more general space-time. While in~\cite{Dzhunushaliev2010}, the
authors numerically discussed some thick brane solutions with pure
gravity. In fact, the authors of~\cite{Dzhunushaliev2010} identified
the contribution of higher order curvature term $f(R)=-\alpha R^n$
($\alpha>0$ and $1<n<2$) with an effective ``matter" source. The
solutions were obtained by analyzing the existence of the fix
points. Besides, the trapping of complex scalar field on the brane
solutions was also discussed in~\cite{Dzhunushaliev2010}.

An ideal thick brane solution should be smooth, stable and possible
to localize gravity and various kinds of matter fields. Whether the
four-dimensional gravity can be reproduced on the brane solutions
in~\cite{Afonso2007,Dzhunushaliev2010} is still unclear. In order to
address the issue of localizing gravity on thick $f(R)$-brane, we
analyze the tensor perturbations for a particular model. Finally, we
apply our results on some of the solutions found
in~\cite{Afonso2007}.

We organize this letter as the follows: In the next section, we set
up our model and give the dynamical equations. In section
\ref{section3} we discuss the tensor perturbations of the model. The
localization of four-dimensional gravity on some thick branes is
investigated in section \ref{section4}. We focus mainly on the
solutions for which the bulk curvatures are constants. For such
solutions the perturbed equations reduce to the Klein-Gordon
equation for massless spin-2 particles. Our summary is given in
section \ref{section5}.

\section{The model}
\label{section2}

We start with the action
\begin{eqnarray}
  S=\int d^4x dy\sqrt {-g}\left(\frac{1}{2\kappa_5^2}f(R)
  -\frac12\partial^M\phi\partial_M\phi-V(\phi)\right),
  \label{action}
\end{eqnarray}
where $\phi$ is a background scalar field which generates the brane.
$V(\phi)$ describes a self-interacting potential for the scalar
field. The gravitational coupling constant $\kappa_5^2=8\pi G_5$
with $G_5$ the five-dimensional Newtonian constant. Indices $M, N
\cdots=0,1,2,3,5$ and $\mu, \nu \cdots=0,1,2,3$ are always applied
to denote the bulk and the brane coordinates, respectively.

For simplicity, we consider the static flat braneworld scenarios
with the metric
\begin{eqnarray}
  ds^2=e^{2A(y)}\eta_{\mu\nu}dx^\mu dx^\nu+dy^2,
  \label{metric}
\end{eqnarray} and the scalar field is assumed to be a function of the extra dimension $y=x^4$, i.e., $\phi=\phi(y)$.
For system (\ref{action})-(\ref{metric}), Einstein equations are
\begin{eqnarray}
\label{EE1}
  f(R)+2f_R\left(4A'^2+A''\right)
  -6f'_RA'-2f''_R=\kappa_5^2(\phi'^2+2V),
\end{eqnarray}
and
\begin{eqnarray}
\label{EE2}
  -8f_R\left(A''+A'^2\right)+8f'_RA'
  -f(R)=\kappa_5^2(\phi'^2-2V),
\end{eqnarray} where the primes represent derivatives with respect to
the coordinate $y$ and $f_R=df(R)/dR$. The equation of motion for
the scalar field is given by
\begin{eqnarray}
 4A'\phi'+\phi''-\frac{\partial V}{\partial\phi}=0.
 \label{Eom}
\end{eqnarray}
Obviously, these equations contain higher-order derivatives of the
coordinates. Usually, it is very hard to solve these equations
analytically, needless to say to analyze the feedbacks of these
equations for the perturbations from both the metric and the scalar
field, because, in general, the perturbed equations are coupling
equations with higher-order derivative terms.

\section{Tensor perturbations}
\label{section3}

We consider the following metric perturbations:
\begin{eqnarray}
ds^2=e^{2A(y)}(\eta_{\mu\nu}+h_{\mu\nu})dx^\mu dx^\nu+dy^2,
\end{eqnarray}
or, in another form
\begin{eqnarray}
g_{MN}=\bar{g}_{MN}+\Delta g_{MN},
\end{eqnarray}
with
\begin{eqnarray}
\bar{g}_{MN}= \left(
  \begin{array}{cc}
    e^{2A}\eta_{\mu\nu} & 0 \\
    0 & 1 \\
  \end{array}
\right),\quad \Delta g_{MN}=\left(
  \begin{array}{cc}
    e^{2A}h_{\mu\nu} & 0 \\
    0 & 0 \\
  \end{array}
\right)
\end{eqnarray}
the background metric and the metric perturbations, respectively.
Here $h_{\mu\nu}=h_{\mu\nu}(x^{\rho},~y)$ depend on all the
coordinates. Obviously, $\Delta g_{5M}=0$, which means we consider
only tensor perturbations. According to the relation
$g^{NP}g_{PM}=\delta_M^{~N}$, one obtains the inverse of $\Delta
g_{MN}$, i.e., $\Delta g^{MN}$. We keep only the first order term,
i.e., $\Delta g^{MN(1)}$ and denote it as
\begin{eqnarray}
\delta g^{MN}=\left(
  \begin{array}{cc}
    -e^{-2A}h^{\mu\nu} & 0 \\
    0 & 0 \\
  \end{array}
\right),
\end{eqnarray} where $h^{\mu\nu}=\eta^{\mu\lambda}\eta^{\nu\rho}h_{\lambda\rho}$ is
raised by $\eta^{\mu\nu}$. We always use $\delta X$ to denote the
first order contribution of the perturbations to an arbitrary
quantity $X$. The perturbation of the scalar field is assumed to be
of first order and is denoted by
$\delta\phi=\tilde{\phi}(x^{\mu},~y)$.

Denoting $a(y)\equiv e^{A(y)}$, we immediately obtain the following
relations:
\begin{eqnarray}
  \delta R_{\mu\nu}&=&-\frac{1}{2}(\square^{(4)} h_{\mu \nu }
               +\partial _{\mu }\partial _{\nu }h
               -\partial _{\nu }\partial _{\sigma }h_{\mu }^{\sigma }
               -\partial _{\mu }\partial _{\sigma }h_{\nu }^{\sigma })
               -2 a a' h_{\mu \nu }'\nonumber\\
               &-&3 h_{\mu \nu } a'^2
               -a h_{\mu \nu } a''
               -\frac{a^2 h_{\mu \nu }''}{2}
               -\frac{a \eta _{\mu \nu } a' h'}{2},\nonumber\\
\delta R_{\mu 5}&=&\frac12\partial_y(\partial_\lambda
h_\mu^\lambda-\partial_\mu h),\quad
  \delta R_{55}=-\frac{1}{2} \left(\frac{2 a' h'}{a}+h''\right),\nonumber\\
\delta R&=&\delta (g^{\mu\nu}R_{\mu\nu})=-\frac{\square^{(4)}
h}{a^2}+\frac{\partial _{\mu }\partial _{\nu } h^{\mu \nu }}{ a^2}
        -\frac{ a' }{a}5h'-h''.
        \label{10}
\end{eqnarray}
Here $\square^{(4)}=\eta^{\mu\nu}\partial_{\mu}\partial_{\nu}$, is
the four-dimensional d'Alembert operator, and $h=\eta^{\mu\nu}h_{\mu
\nu}$ is the trace of the tensor perturbations.

Obviously, if
\begin{eqnarray}
h=0=\partial_\mu h^\mu_{~\nu}, \label{TTcondision}
\end{eqnarray} only $\delta
R_{\mu\nu}$ remains non-zero. The condition (\ref{TTcondision}) is
called the transverse-traceless gauge which can largely simplify the
perturbed equations.

  In $f(R)$ gravity, the Einstein equations
are
\begin{eqnarray}
  R_{MN}f_R-\frac12g_{MN}f(R)+(g_{MN}\square^{(5)}-\nabla_M\nabla_N)f_R=\kappa_5^2T_{MN},
  \label{eqEE}
\end{eqnarray} with $\square^{(5)}=g^{MN}\nabla_{M}\nabla_{N}$ the five-dimensional d'Alembert operator.
As the perturbations are considered, the feedback from Einstein
equations reads
\begin{eqnarray}
 &&\delta R_{MN}f_R+R_{MN}f_{RR}\delta R-\frac12\delta g_{MN}f(R)
 -\frac12g_{MN}f_R\delta R\nonumber\\
 &+&\delta(g_{MN}\square^{(5)} f_R)-\delta(\nabla_M\nabla_Nf_R)=\kappa_5^2\delta
 T_{MN}.
 \label{eqPertubedEE}
\end{eqnarray}
We need only to calculate the second line of the above equation.
Note that
\begin{eqnarray}
\nabla_M\nabla_Nf_R&=&(\partial_M\partial_N-\Gamma^P_{MN}\partial_P)f_R,\nonumber\\
g_{MN}\square^{(5)} f_R&=&g_{MN}g^{AB}(\nabla_A\nabla_B f_R),
\end{eqnarray} we have
\begin{eqnarray}
  \delta(\nabla_M\nabla_Nf_R)&=&(\partial_M\partial_N-\Gamma^P_{MN}\partial_P)(f_{RR}\delta
  R)-\delta \Gamma^P_{MN}\partial_P f_R,\\
  \delta(g_{MN}\square^{(5)} f_R)&=&\delta g_{MN}\square^{(5)} f_R
  +g_{MN} \delta g^{AB}(\nabla_A\nabla_B f_R)\nonumber\\
  &+&g_{MN}g^{AB}\delta(\nabla_A\nabla_Bf_R).\label{34}
\end{eqnarray}
The fluctuations of the energy-momentum
\begin{eqnarray}
 T_{MN}
    =\nabla_M\phi\nabla_N\phi-\frac12g_{MN}g^{AB}\nabla_A\phi\nabla_B\phi
    -g_{MN}V(\phi)
\end{eqnarray} are given by
\begin{eqnarray}
 \delta T_{\mu\nu}&=&-\frac{a^2}{2}\phi'^2h_{\mu\nu}
                         -a^2\eta_{\mu\nu}\phi'\tilde{\phi}'
                         -a^2Vh_{\mu\nu}-a^2\eta_{\mu\nu}\frac{\partial V}{\partial\phi}\tilde{\phi},\nonumber\\
 \delta T_{5\mu}&=&\phi'\partial_\mu \tilde{\phi}, \quad \delta T_{55}=\phi'\tilde{\phi}'-\frac{\partial
 V}{\partial\phi}\tilde{\phi}.
 \label{18}
 \end{eqnarray}
Note that under the transverse and traceless gauge, $\delta R=0$,
and eqs.~(\ref{34}) reduce to
\begin{eqnarray}
  \delta(\nabla_M\nabla_Nf_R)&=&-\delta^\rho_M\delta^\sigma_N\delta\Gamma^5_{\rho\sigma} f_R'~,\nonumber\\
  \delta(g_{MN}\square^{(5)} f_R)&=&\delta^\rho_M\delta^\sigma_N\delta g_{\rho\sigma} \square^{(5)}
f_R.
\end{eqnarray} Therefore, under this gauge, we have
\begin{eqnarray}
  &&\delta(g_{MN}\square^{(5)} f_R)-\delta(\nabla_M\nabla_Nf_R)\nonumber\\
&=&\delta^\rho_M\delta^\sigma_Na^{2}\left[h_{\rho\sigma}\left(3\frac{a'}{a}f'_R+f''_R\right)
   -\frac12f_R'h_{\rho\sigma}'\right].
   \label{20}
\end{eqnarray}
The perturbed Einstein equations (\ref{eqPertubedEE}) reduce to
\begin{eqnarray}
  &&\delta R_{MN}f_R-\frac12\delta g_{MN}f(R)\nonumber\\
 &+&\delta^\rho_M\delta^\sigma_N a^{2}\left[h_{\rho\sigma}\left(3\frac{a'}{a}f'_R+f''_R\right)
   -\frac12h_{\rho\sigma}'f_R'\right]=\kappa_5^2\delta T_{MN}.
   \label{21}
\end{eqnarray}
By plugging (\ref{10}) and (\ref{18}) into (\ref{21}), we obtain the
$(\mu,\nu)$ components of the perturbed Einstein equations
\begin{eqnarray}
  &&\left(-\frac{1}{2}\square^{(4)} h_{\mu \nu }
               -3 h_{\mu \nu }a'^2
               -2 a a' h_{\mu \nu }'
               -a h_{\mu \nu } a''
               -\frac{a^2 h_{\mu \nu }''}{2 }\right)f_R\nonumber\\
               &&-\frac12a^2h_{\mu\nu}f(R)
               +a^{2}\left[h_{\mu\nu}\left(3\frac{a'}{a}f'_R+f''_R\right)
   -\frac12h_{\mu\nu}'f_R'\right]\nonumber\\
   &&=\kappa_5^2\left(-\frac{a^2}{2}\phi'^2h_{\mu\nu}
  -a^2\eta_{\mu\nu}\phi'\tilde{\phi}'
  -a^2Vh_{\mu\nu}-a^2\eta_{\mu\nu}\frac{\partial
  V}{\partial\phi}\tilde{\phi}\right).
  \label{A38}
\end{eqnarray}
Note that the $(\mu,~\mu)$ components of Einstein equations
(\ref{eqEE}) is
\begin{eqnarray}
  f(R)+2f_R\left[3\left(\frac{a'}{a}\right)^2+\frac{a''}{a}\right]
  -6f'_R\frac{a'}{a}-2f''_R=\kappa_5^2(\phi'^2+2V),
  \label{A39}
\end{eqnarray} which is exactly Eq.~(\ref{EE1}).
We can simplify Eq.~(\ref{A38}) as
\begin{eqnarray}
  &&\left(-\frac{1}{2}\square^{(4)} h_{\mu \nu }
               -2 a a' h_{\mu \nu }'
               -\frac{a^2 h_{\mu \nu }''}{2 }\right)f_R
               -\frac12a^{2}h_{\mu\nu}'f_R'\nonumber\\
   &&=\kappa_5^2\left(-a^2\eta_{\mu\nu}\phi'\tilde{\phi}'
   -a^2\eta_{\mu\nu}\frac{\partial V}{\partial\phi}\tilde{\phi}\right).
\end{eqnarray}
 Contracting the above equation with $\eta^{\mu\nu}$
one can proof that $\phi'\tilde{\phi}'+\frac{\partial
V}{\partial\phi}\tilde{\phi}=0$. Therefore, the $(\mu,~\nu)$
components of the perturbed Einstein equations read
\begin{eqnarray}
  &&\left(a^{-2}\square^{(4)} h_{\mu \nu }
               +4 \frac {a'}{a} h_{\mu \nu }'
               + h_{\mu \nu }''
               \right)f_R+h_{\mu\nu}'f_R'=0,
\end{eqnarray}
or, equivalently,
\begin{eqnarray}
\square^{(5)}h_{\mu\nu}=\frac{f_R'}{f_R}\partial_y h_{\mu\nu}.
\label{eqEEFinal}
\end{eqnarray}
With the coordinate transformation
\begin{eqnarray}
dz=a^{-1}dy,\label{coordinatetrans}
\end{eqnarray}
we can rewrite the perturbed equation (\ref{eqEEFinal}) as
\begin{eqnarray}
 \left[\partial_z^{~2}
  +\left(3\frac{\partial_z a}{a}
  +\frac{\partial_z f_R}{f_R}\right)\partial_z
  +\square^{(4)}\right]h_{\mu\nu}=0.
\end{eqnarray}
Consider the decomposition
$h_{\mu\nu}(x^{\rho},z)=(a^{-3/2}f_R^{-1/2})\epsilon_{\mu\nu}(x^{\rho})\psi(z)$,
and ask $\epsilon_{\mu\nu}(x^{\rho})$ satisfies the transverse and
traceless condition $\eta^{\mu\nu}\epsilon_{\mu\nu}=0=\partial_\mu
\epsilon^{~\mu}_\nu$, we would have a Schr\"odinger equation for
$\psi(z)$:
\begin{eqnarray}
  \left[\partial_z^2
      -W(z)\right]\psi(z)
      =-m^2\psi(z),\label{Schrodinger}
\end{eqnarray} with the potential $W(z)$ given by
\begin{eqnarray}
 W(z)=\frac34\frac{a'^2}{a^2}
      +\frac32\frac{a''}{a}
      +\frac32\frac{a' f_R'}{a f_R}
      -\frac14\frac{f_R'^2}{f_R^2}
      +\frac12\frac{f_R''}{f_R}.
      \label{Schrodingerpotential}
\end{eqnarray}
To be more explicit, one can also factorize the Schr\"odinger
equation (\ref{Schrodinger}) as
\begin{eqnarray}
 \left[\left(\partial _z
 +\left(\frac{3}{2}\frac{\partial_z a}{a}+\frac{1}{2}\frac{\partial_z f_R}{f_R}\right)\right)
 \left(\partial_z
 -\left(\frac{3}{2}\frac{\partial_z a}{a}+\frac{1}{2}\frac{\partial_z f_R}{f_R}\right)\right)\right]\psi(z)
 =-m^2\psi(z),
\end{eqnarray} which indicates that there
is no gravitational mode with $m^2<0$. Therefore any solution of the
system (\ref{action})-(\ref{metric}) is stable under the tensor
perturbations. The zero mode (if exists) takes the form
\begin{eqnarray}
\psi^{(0)}(z)=N_0 a^{3/2}(z)f_R^{1/2}(z), \label{zeromode}
\end{eqnarray} with $N_0$ the normalization constant.

These results indicate that as the transverse and traceless gauge is
taken, the perturbation of the scalar field decouples from the
metric perturbations. In the case of general relativity,
Eq.~(\ref{eqEEFinal}) reduce to the five-dimensional Klein-Gorden
equation for the massless spin-2 gravitons. However, for an
arbitrary form of $f(R)$ and non-constant curvature $R$, the
equation for $h_{\mu\nu}$ is largely different from the massless
Klein-Gorden equation. Fortunately, the perturbed equation always
remains second order due to the introducing of the transverse and
traceless gauge. Now let us see a simple application of our results
on some solutions given previously in~\cite{Afonso2007}.

\section{Applications}
\label{section4}

In~\cite{Afonso2007}, the authors gave us brane solutions for both
constant and variant curvature cases. However, the solution for
later case contains a singular point, and therefore is not
regularized. So we would like to consider only the constant
curvature case. The corresponding Einstein equations in this case
reduce to some second ones:
\begin{subequations}
\label{consEE}
  \begin{equation}
    f(R)+2f_R\left(3\left(\frac{a'}{a}\right)^2+\frac{a''}{a}\right)
  =\kappa_5^2(\phi'^2+2V),
  \end{equation}
  \begin{equation}
    -8f_R\frac{a''}{a}
  -f(R)=\kappa_5^2(\phi'^2-2V).
  \end{equation}
\end{subequations}
Constraining $R$ as a constant, the solution for warp factor is
uniquely determined~\cite{Afonso2007}:
\begin{subnumcases}
 {a(y)=}
   (\frac{5}{2}k y)^{2/5},
              & for $R=0$, \label{R0}\\
   \cos^{2/5}\left(\frac{5}{2} ky\right),
              & for $R=20k^2>0$, \label{R+}\\
   \cosh^{2/5}\left(\frac{5}{2} ky\right),& for $R=-20k^2<0$.
 \label{R-}
\end{subnumcases}
Note that the warp factor given in (\ref{R0}) is not smooth, it
contains a cusp at $y=0$. This cusp of the warp factor leads to at
most a $\delta$-function in the second-order Einstein gravity, which
can be explained as the appearance of a thin brane. However, in the
fourth-order $f(R)$ gravity, such cusp will introduce derivatives of
$\delta(y)$, such as $\delta'(y)$ and $\delta''(y)$, and cross terms
of them. It is still a problem of how to deal with such terms in
brane theory. For this reason, we consider only solutions for $dS_5$
space where $R>0$; and for $AdS_5$ space where $R$ is a negative
constant.

One can easily proof that for $dS_5$ and $AdS_5$ spaces,
eqs.~(\ref{consEE}) support the following non-trivial solutions:
\begin{itemize}
 \item For $dS_5$ and $f_R>0$
\begin{eqnarray}
 \phi& =&\pm \sqrt{\frac{6f_R}{5\kappa _5^2}}\textrm{arctanh}(\sin(5k
 y/2)),\\
  V(\phi)&=&V_1-\frac{9 k^2f_R}{4 \kappa _5^2} \sinh^2 \left(\sqrt{\frac{5\kappa _5^2}{6f_R}}\phi\right),
\end{eqnarray}
 with $V_1=\frac{2 f(R)-25 k^2 f_R}{4 \kappa
_5^2}$. We consider the interval $-\frac{\pi}{5k} \leq y \leq
\frac{\pi}{5k}$, within which the warp factor (\ref{R+}) is
regularized.
  \item For $AdS_5$, and $f_R<0$
\begin{eqnarray}
\label{611}
\phi&=&\pm \sqrt{\frac{6|f_R|}{5\kappa _5^2}}\textrm{arctan}\left(\sinh(5k y/2)\right),\\
V(\phi)&=&V_2+\frac{9 k^2|f_R|}{4 \kappa _5^2}\sin^2
\left(\sqrt{\frac{5\kappa _5^2}{6|f_R|}}\phi\right),
\end{eqnarray} with $V_2\equiv\frac{f(R)}{2 \kappa _5^2}-\frac{25 k^2 |f_R|}{4 \kappa _5^2}$.
\end{itemize}

\begin{figure}
\includegraphics[width=0.5\textwidth]{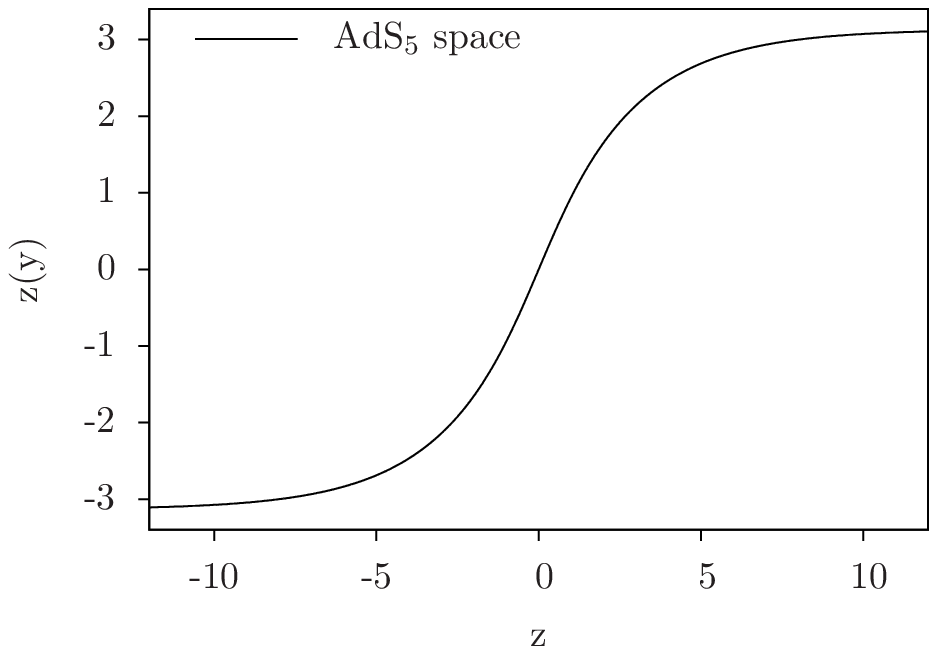}
\includegraphics[width=0.5\textwidth]{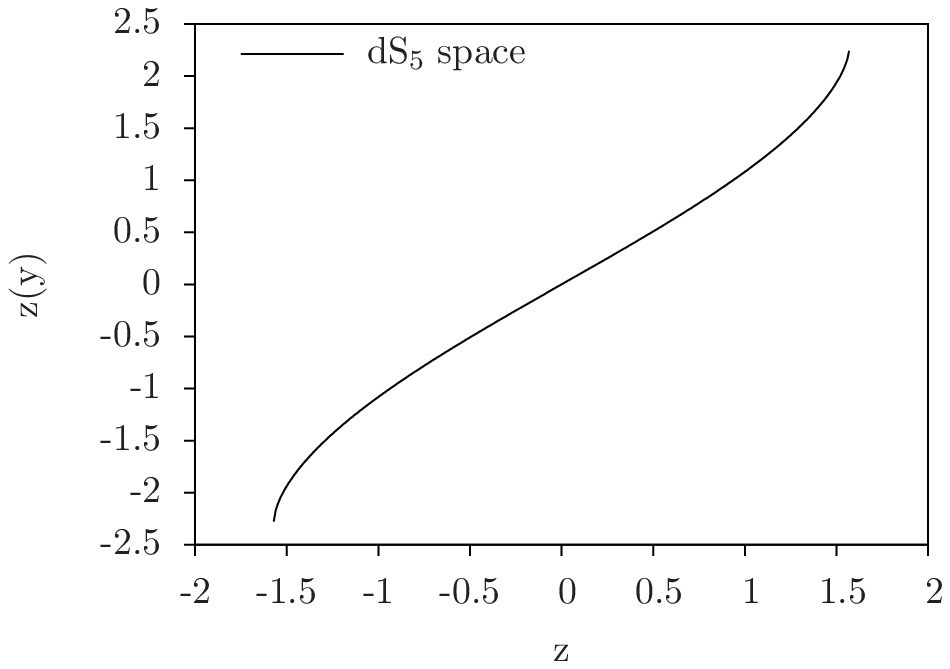}
\caption{Figure of $z(y)$ for $AdS_5$ space (left panel) and for
$dS_5$ space (right panel) with $k=2/5$.} \label{figurezy}
\end{figure}

For the warp factors given in (\ref{R+}) and (\ref{R-}), the
coordinate transformation (\ref{coordinatetrans}) cannot be
integrated out analytically. However, we can find the numerical
relation between $z$ and $y$, see figure~\ref{figurezy}. Note that
for the case of $dS_5$ space,
\begin{eqnarray}
z(y)\leq \int_{0}^{\frac{\pi}{5k}}
\cos^{-\frac{2}{5}}\big(\frac{5}{2}ky\big)dy
 =\frac{\sqrt{\pi }\Gamma\left(\frac{3}{10}\right)}{5k\Gamma\left(\frac{4}{5}\right)},
\end{eqnarray}
where $\Gamma (\beta)=\int _0^{\infty }t^{\beta-1}e^{-t}dt$ is the
Euler gamma function. Therefore, $z(y)$ is a bounded function of
$y$.

For constant curvature spaces $f_R'=0$, the perturbed Einstein
equations reduce to
\begin{eqnarray}
\square^{(5)}h_{\mu\nu}=0.
\end{eqnarray}
This is the familiar Klein-Gordon equation one obtains in general
relativity~\cite{Csaki2000a,DeWolfe2000a}. To analyze the
possibility of localizing four-dimensional gravity on the brane, we
simply solve the shr\"odinger equation~(\ref{Schrodinger}) with the
following potential:
\begin{eqnarray}
 W(z)=\frac34\frac{a'^2}{a^2}
      +\frac32\frac{a''}{a}.
\end{eqnarray} The corresponding zero mode takes the form
\begin{eqnarray}
\psi^{(0)}(z)\propto a^{3/2}(z).
\end{eqnarray}

\begin{figure}
\includegraphics[width=0.5\textwidth]{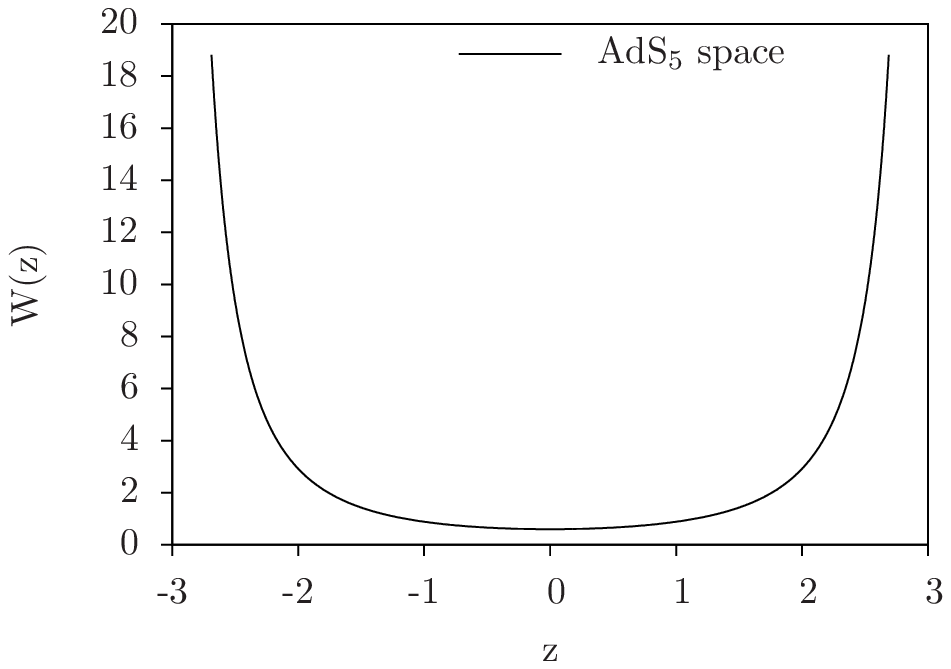}
\includegraphics[width=0.5\textwidth]{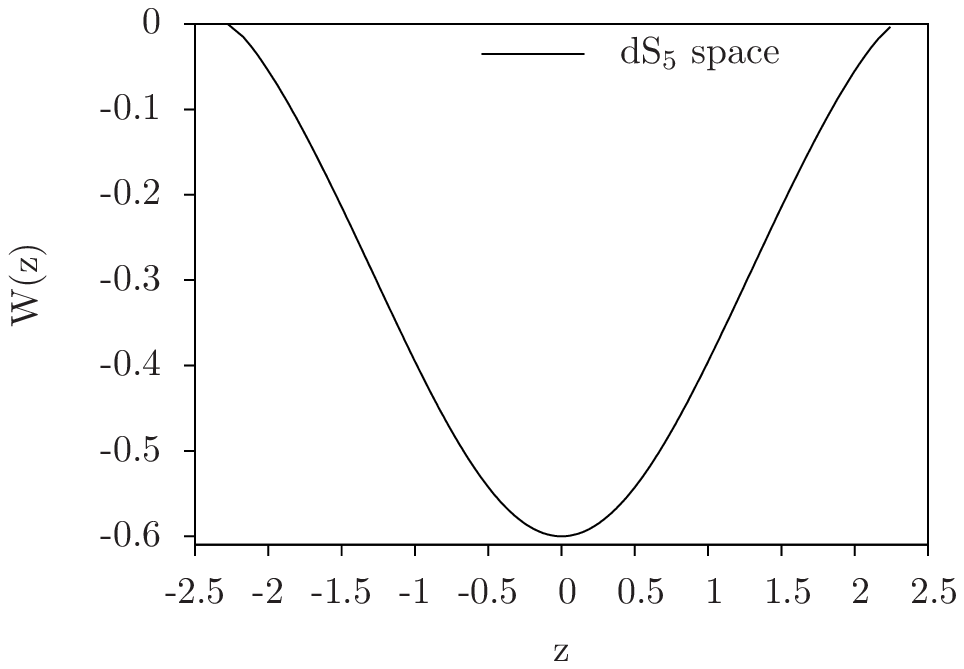}
\caption{Figure of $W(z)$ for $AdS_5$ space (left panel) and for
$dS_5$ space (right panel) with $k=2/5$.} \label{figureWz}
\end{figure}

As shown in figure~\ref{figureWz}, for the case of $AdS_5$
space-time, the potential $W(z)$ is positive everywhere and diverge
at $z=\pm\infty$. Such potential supports only bounded and discrete
KK states. There is no continuum spectrum. Thus, the solution itself
is stable under tensor perturbations. However, the zero mode dose
not exist, as a consequence, the four-dimensional massless graviton
cannot be localized on the brane.

%

\begin{figure}[h]
\begin{center}
\includegraphics[width=0.5\textwidth]{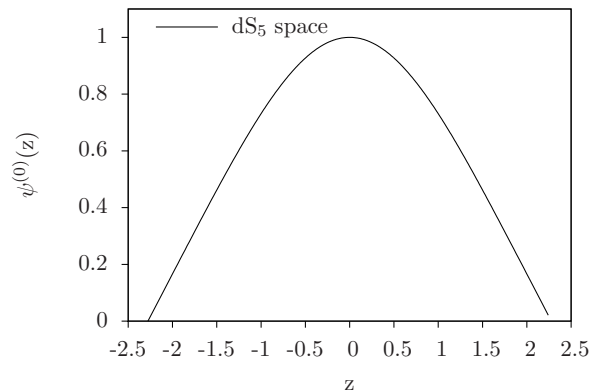}
\caption{The zero mode $\psi^{(0)}(z)$ for the case of $dS_5$ space
with $k=2/5$.}
 \label{zeromode}
 \end{center}
\end{figure}
The potential $W(z)$ for $dS_5$ space, as shown in
figure~\ref{figureWz}, is a bounded function. 
Both the potential $W(z)$ and the zero mode (see
figre~\ref{zeromode}) vanish at the boundary
$z(y=\pm\frac{\pi}{5k})=\pm\frac{\sqrt{\pi }
\Gamma\left(\frac{3}{10}\right)}{5k\Gamma\left(\frac{4}{5}\right)}$.
That means the solution in the case of $dS_5$ is stable; in
addition, the four-dimensional graviton can be localized on the
brane.

\section{Conclusions}
\label{section5}

To sum up, we considered the fluctuations from both the metric and
the scalar field around the flat thick $f(R)$-branes. It turns out
that the perturbation from the scalar field decouples from the
tensor part of the metric perturbations when the transverse and
traceless gauge is considered. The propagation of the metric
perturbations in the bulk is not described by the simple massless
Klein-Gordon equation any more, except $f(R)=R$ or $R$ is a
constant. As an application of our results, we studied the stability
of some simple solutions given previously. These solutions were
found by constraining the bulk curvatures as constants. Among these
solutions, the one for $R=0$ is problematic because at the location
of the brane the metric poses a cusp, which would lead to the
problem of divergence in the fourth-order $f(R)$ gravity. The
analysis of the solution in case of $AdS_5$ indicates that the
solution is stable, there are infinite discrete massive KK states.
While for the case of $dS_5$, the solution is stable, and the
normalizable zero mode does exist. Therefore, the four-dimensional
massless graviton can be localized on $dS_5$ brane.

The results we obtained in this letter are valid for any solution of
the system (\ref{action})-(\ref{metric}). In the present letter, we
confined our discussions only on the simplest case, i.e.,
$R=$const.. However, the constant curvature spaces are rather
special, because in this case the equations of motion of the tensor
perturbations are independent of the form of $f(R)$. Thus it is
natural to ask what would be different if the curvature is variant.
To answer this question we have to firstly find a well behaved thick
$f(R)$-brane solution which is regularized, stable, and analytical
(rather than numerical\footnote{In fact, some numerical thick
$f(R)$-brane solutions have been found in~\cite{Dzhunushaliev2010},
where $R$ is a function of the fifth dimension $y$. However, it is
very hard for us to analyze these numerical solutions.}). We also
hope that the localization of gravity and the trapping of bulk
matters are guaranteed by these solutions. Unfortunately, there is,
to our knowledge, no such solution has ever been reported. In one of
our recent works \cite{ZhongfR2}, we have found that at least in
squared curvature gravity (where $f(R)\propto R^2$), there exists a
thick domain wall solution which possess nearly all the properties
we are searching for. It seems that for the case of variant
curvature, the thick $f(R)$-brane solutions contain more interesting
features.

\chapter{\textbf{Acknowledgments}}

The authors would like to thank the anonymous referees whose
comments largely helped us in improving the original manuscript.
This work was supported by the Program for New Century Excellent
Talents in University, the Huo Ying-Dong Education Foundation of
Chinese Ministry of Education (No. 121106), the National Natural
Science Foundation of China (No. 11075065), the Doctoral Program
Foundation of Institutions of Higher Education of China (No.
20090211110028), and the Natural Science Foundation of Gansu
Province, China (No. 096RJZA055).



\end{document}